\documentclass[prl,twocolumn]{revtex4-1}
\usepackage{bm}
\usepackage{graphicx}
\usepackage{amssymb}
\usepackage{amsmath}
\usepackage{eufrak}
\usepackage{color}
\usepackage{pifont}
\usepackage{array}
\usepackage{verbatim} 
\usepackage{placeins}
\usepackage{float}
\usepackage[T1]{fontenc}
\usepackage{color}
\usepackage[colorlinks]{hyperref}
\usepackage[normalem]{ulem}

\newcommand{\eps}{\varepsilon}

\begin{document}
	\title{Edge currents driven by terahertz radiation in graphene in quantum Hall regime}
	\author{ Plank~H$^1$, Durnev~M~V$^2$, Candussio~S$^1$, Pernul~J$^1$, Dantscher~K-M$^1$, M\"onch~E$^1$, Sandner~A$^1$, Eroms~J$^1$, Weiss~D$^1$, Bel'kov~V~V$^2$, Tarasenko~S~A$^2$ and Ganichev~S~D$^1$}	
	\affiliation{$^1$~Terahertz Center, University of Regensburg, 93053 Regensburg, Germany}
	\affiliation{$^2$~Ioffe Insitute, 194021 St. Petersburg, Russia}
	\begin{abstract}	
		We observe that the illumination of \textit{unbiased} graphene in the quantum Hall regime with polarized terahertz laser radiation results in a direct edge current. 
		This photocurrent is caused by an imbalance of persistent edge currents, which are driven out of thermal equilibrium by indirect transitions within the chiral edge channel. 
		The direction of the edge photocurrent is determined by the polarity of the external magnetic field, while its magnitude depends on the radiation polarization. 
		The microscopic theory developed in this paper  describes well the experimental data.
	\end{abstract}
	\maketitle

	\section{Introduction} 
	Conducting channels emerging at the edges of materials with non-trivial topology, e.g., in the quantum Hall regime or in two-dimensional (2D) topological insulators, is at the core of the physics of 2D systems~\cite{Hasan2010,Qi2011}. 
	One-dimensional chiral channels with persistent charge currents circulating around the sample's edges naturally occur in 2D electron systems subjected to a quantizing perpendicular magnetic field, i.e., at quantum Hall conditions~\cite{Halperin1982,Buettiker1988,Altimiras2010,Koenig2007}. 
	Graphene with its Dirac-like electron spectrum is a unique system for the study of edge channel effects because the large cyclotron gap, essential for the formation of edge channels, of 30 meV is achieved already in the magnetic field below 1~T~\cite{Novoselov2005,Zhang2005,Novoselov2007,Tzalenchuk2010, Young2014,Ribeiro-Palau2015}. 
	So far, the chiral edge transport of carriers in graphene was studied in the ballistic regime with suppressed scattering when the current is driven by an applied voltage.
	
	Here, on example of unbiased graphene, we show that a direct electric current in chiral edge channels can be excited by terahertz radiation with photon energies smaller than the cyclotron gap. The direction of this edge photocurrent is determined by the polarity of the external magnetic field while the radiation polarization affects only its amplitude. 
	Such a photocurrent is demonstrated to be caused by unbalancing persistent currents when driving the system out of thermal equilibrium. 
	The mechanism of the photocurrent generation is microscopically strikingly different from the previously reported 
	magnetic-field-induced photoelectric effects in graphene structures which rely on the 2D motion of free carriers or inter-band optical transitions. 
	Such photocurrents generated due to the presence 
	of $dc$ or $ac$ magnetic fields include magnetic quantum ratchets in structures with broken space inversion symmetry~\cite{Drexler2014,Belkov2005,Tarasenko2011,Kheirabadi2016,Kheirabadi2018}, cyclotron resonance assisted edge photocurrents excited by infrared radiation~\cite{Masubuchi2013,Sonntag2017}, dynamic Hall effect and photon drag~\cite{karch2010,entin10,jiangPRB2011,Obraztsov2014}, 
	etc., for recent reviews see~\cite{GlazovGanichev_review,Koppens14,Ganichev2017}.
	Analyzing the magnitude of the chiral edge currents observed in the present work we find that the relaxation time of non-equilibrium carriers in chiral edge channels is two orders of magnitude longer than the momentum relaxation time of bulk carriers at zero magnetic field.
	
	\section{Methods}
\begin{figure}
		\includegraphics[width=\linewidth]{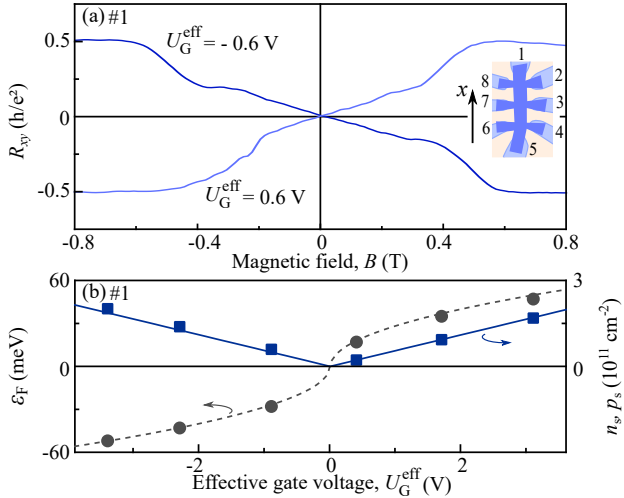}
		\caption{ 
			Magneto-transport results obtained for sample \#1. 
			(a) Transverse resistance $R_{xy}$ for two different gate voltages corresponding to $n$- and $p$- type conductivity. 
			The inset shows the Hall bar geometry and the corresponding contact numbers. 
			(b) Carrier densities $n_s, p_s$ and Fermi energy $\varepsilon_{\rm F}$ as a function of the effective back gate voltage $U^{\rm eff}_{\rm G}$. 
		}
		\label{transport}
	\end{figure}

	Experiments are performed on Hall bar structures prepared from exfoliated graphene/hexagonal boron nitride stacks~\cite{Dean2010,Wang2013,Sandner2015}, see inset figure\,\ref{transport}(a). 
	At zero back gate voltage the samples used have mobilities and hole densities of 
	$\mu = 9.9 \times 10^4$~cm$^2$/Vs, $p_s = 4 \times 10^{10}$~cm$^{-2}$ (sample \#1) and 
	$\mu = 3.5 \times 10^4$~cm$^2$/Vs, $p_s = 8 \times 10^{10}$~cm$^{-2}$ (sample \#2) at $T = 4.2$~K. 
	Varying the back gate voltage $U_{\rm G}$, we can tune the Fermi level with the charge neutrality point (CNP) at a voltage of about 1~V. 
	Note that for different sample cool-downs, the CNP position $U_{\rm CNP}$ is shifted slightly. 
	Therefore, to compare various measurements we use the normalized gate voltage $U^{\rm eff}_{\rm G} = U_{\rm G} - U_{\rm CNP}$.
	Figure\,\ref{transport}(a) presents magneto-transport measurements for two effective gate voltages corresponding to $p$- and $n$-type conductivity. 
	The data show that 
	the quantum Hall plateau with filling factor 2 is reached below 0.8~T. 
	The variation of the carrier density and Fermi energy as a function of the effective gate voltage is plotted in figure\,\ref{transport}(b). 
	
	Edge currents are excited with normally incident terahertz radiation of pulsed molecular laser~\cite{schneider2004,lechner2009} using NH$_3$, D$_2$O and CH$_3$F as active media and operating at frequencies $f = 0.6$, 0.8, 1.1, 2.0, and 3.3~THz. 
	The gas laser is pumped by transversely excited atmospheric pressure (TEA) CO$_2$ laser and provides pulses with width in the order of 100~ns and peak pulse powers in the order of tens of kW. The peak power of the radiation was monitored with photon-drag detectors~\cite{Ganichev84p20}.
	The spot of the terahertz radiation, measured with pyroelectric camera, has an almost Gaussian profile and is about 1.5 - 3~mm in diameter. 
	Consequently, the micrometer sized Hall bar structures are illuminated homogeneously with radiation intensity of 5 - 200~kW/cm$^2$ depending on the wavelength.
	The orientation of the radiation electric field vector ${\bm E}(t)$ is controlled by crystal quartz $\lambda$/2 wave plates and is defined by the azimuthal angle $\alpha'$, 
	with $\alpha'$ being the angle between the electric field of the radiation and the sample edge. By that for $\alpha' = 0$ ${\bm E}(t)$ is parallel to $x$ . 
	An external magnetic field up to $\pm 2$~T is applied normal to the sample surface and parallel to the radiation propagation. 
	Photocurrents are measured with a digital oscilloscope as a voltage drop across a 50~$\Omega$ load resistor in samples, which are cooled down to $T = 4.2$~K. 
		
	\section{Results}
	\subsection{Edge currents at zero magnetic field}
\begin{figure}
		\includegraphics[width=\linewidth]{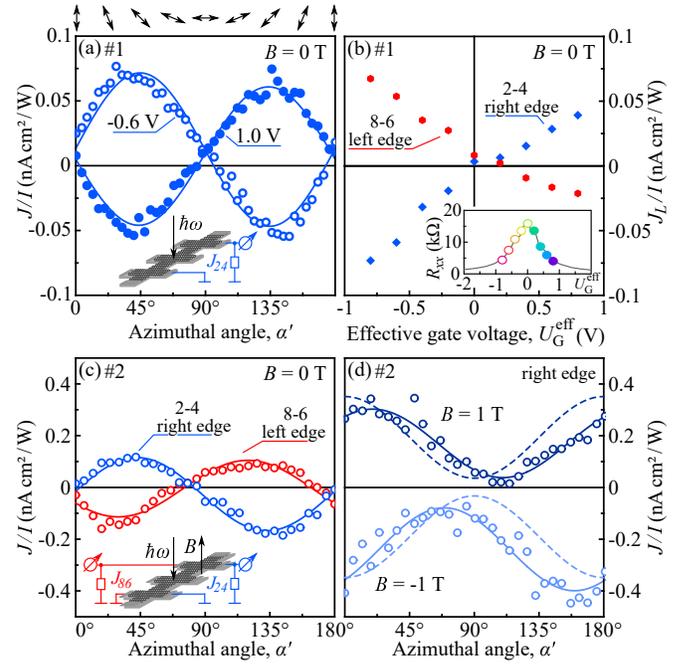}
		\caption{ 
			Photocurrents excited at graphene edges by radiation with $f=3.3$~THz. 
			(a) Azimuthal angle dependence of the normalized photocurrent measured in sample \#1 	
			between contacts 2-4 (right edge) at zero magnetic field. 
			Data are presented for two effective gate voltages $U^{\rm eff}_{\rm G}$	corresponding to $n$- (solid circles) and $p$- (open circles) type of conductivity.
			Curves are fits after \eqref{fit-alpha}. 
			Inset shows experimental setup. 
			Arrows on top of both panels illustrate states of polarization for several angles $\alpha'$. 
			(b) Amplitude of the photocurrent $J_L$ as a function of the effective gate voltages $U^{\rm eff}_{\rm G}$ obtained for opposite edges: 
			left edge (contacts 8-6) and the right one (contacts 2-4). 
			The inset shows the longitudinal resistance $R_{xx}$ as a function of the effective gate voltage $U^{\rm eff}_{\rm G}$. 
			Colored solid and open circles indicate gate voltages used in the photocurrent measurements presented in figure~\protect\ref{alpha-gate-B}. 
			(c) Azimuthal angle dependence of the normalized photocurrent $J/I$ measured in sample \#2 
			for opposite edges. Curves are fits after \eqref{fit-alpha}. Inset shows experimental setup. 
			(d) Polarization dependence of the photocurrent measured for the right edge 
			at magnetic field $B = \pm 1$~T, corresponding to the Hall plateau with filling factor~2. Solid curves are fits after \eqref{fit-alpha}, dashed curves are calculated after \eqref{jedge2} and \eqref{eta}. 
		}		
		\label{alpha}
	\end{figure}

	A photocurrent excited by polarized radiation is detected between different contact pairs along the long Hall bar edges for all the frequencies used. 
	Figure\,\ref{alpha}(a) shows a typical dependence of the photocurrent at zero magnetic field on the orientation of the electric field vector picked up at the right edge of sample \#1 (contacts 2-4), see the inset in this panel. 
	The strength and direction of the photocurrent are controlled by the radiation polarization varying with the azimuthal angle $\alpha'$ as 
	\begin{equation} \label{fit-alpha}
	J = J_{\rm L} \sin(2\alpha' + \psi) + J_{\rm 0}, 
	\end{equation}
	where $J_{\rm L}$ and $J_{\rm 0}$ are the amplitudes of the polarization dependent and independent contributions, respectively. 
	Note that at zero magnetic field, $J_{\rm 0}$ is much smaller than $J_{\rm L}$ and the phase $\psi$ is close to zero. 
	Changing the carrier type from electrons to holes reverses the direction of the current, see figure\,\ref{alpha}(a). 
	The variation of the amplitude $J_{\rm L}$ with the carrier density is shown in figure\,\ref{alpha}(b), in which the dependence on the effective gate voltage $U^{\rm eff}_{\rm G}$ is presented for two opposite edges. The inset in this figure 
	shows the corresponding dependence of the longitudinal resistance.
	This figure and figure\,\ref{alpha}(c) reveal that the photoresponse is consistently of opposite signs for the opposite edges and, therefore, originates from edge photocurrents. 
	Similar results were obtained previously in large area low mobility samples~\cite{GlazovGanichev_review,edge}. The origin of the edge photocurrent at zero magnetic field will be briefly addressed below.
	
	\subsection{Edge currents in the quantum Hall regime}
\begin{figure}
		\includegraphics[width=\linewidth]{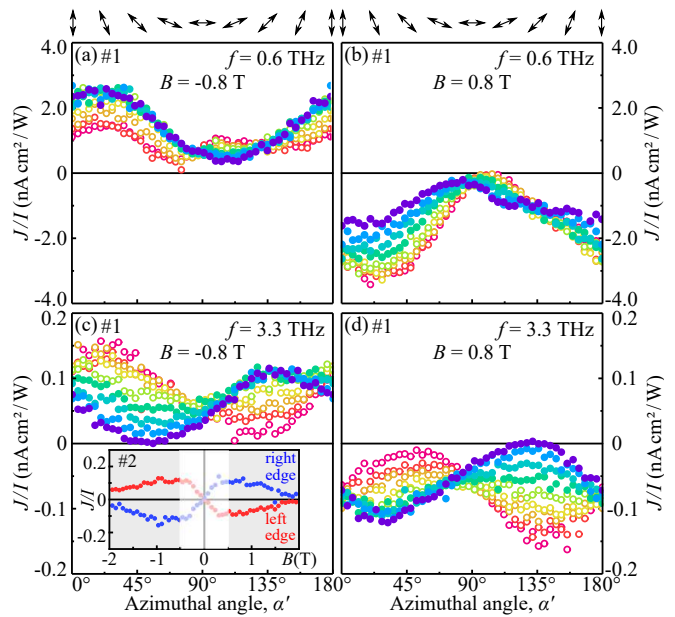}
		\caption{
			Dependence of the normalized photocurrent on the orientation of the radiation electric field vector, obtained for the right edge of the sample \#1. 	
			Data are shown for magnetic fields $B = \pm 0.8$~T and gate voltages $U^{\rm eff}_{\rm G}$ in the range from - 0.8 to 1~V. 
			Panels (a) and (b) show data for frequency $f = 0.6$~THz, and panel (c) and (d) for $f = 3.3$~THz. 
			Solid and open circles correspond to electrons and holes, respectively. 
			Circle colors indicate the values of the effective gate voltage $U^{\rm eff}_{\rm G}$, which are shown in the inset in figure\,\ref{alpha}(b). 
			Inset in panel (c) shows magnetic field dependence of the photocurrent measured at right and left edges. 
			Data are obtained with $\alpha' = 75^{\circ}$ in sample \#2 at excitation frequency $f = 3.3$~THz. 
			For $|B| > 0.5$~T the sample is in the quantum Hall regime, highlighted by a gray background. 
		}
		\label{alpha-gate-B} 
	\end{figure}

	Driving the system into the quantum Hall regime by application of a magnetic field we observed that the photocurrent behavior changes qualitatively. 
	Now, in contrast to the discussed case of zero magnetic field, see figure\,\ref{alpha}(c), the photocurrent does not change its sign upon variation of the electric field vector orientation, see figure\,\ref{alpha}(d). 
	Its direction is defined by the magnetic field polarity. 
	It is, however, still consistently opposite for opposite edges confirming that the photoresponse is caused by the edge currents, see inset in figure\,\ref{alpha-gate-B}(c). 
	To support this statement we grounded all remaining contacts and observed that this change in the electric circuit neither affects the amplitude nor the polarization dependence of the photocurrents measured (not shown). 
	
	Our measurements reveal that for all the frequencies and gate voltages used, the polarization behavior of the photocurrent in the quantum Hall regime is still described by \eqref{fit-alpha}, but we obtain a substantial contribution of $J_{\rm 0}$ close to $J_L$ and a phase angle $\psi$ close to $90^\circ$ as compared to the data at $B = 0$. 
	Furthermore and in contrast to the zero magnetic field results, switching from $n-$ to $p-$type conductivity does not result in the inversion of the edge photocurrent direction.
	This is shown in figure\,\ref{alpha-gate-B} for different carrier densities $n_s$ (solid circles) and $p_s$ (open circles) and two radiation frequencies $f = 0.6$~THz, in panels (a) and (b), and $f = 3.3$~THz, in panels (c) and (d). 
	Here, the carrier densities obtained from the Hall slopes (not shown) vary from $p_s = 3 \times 10^{10}$~cm$^{-2}$ to $n_s= 7 \times 10^{10}$~cm$^{-2}$. 
	The corresponding longitudinal resistance in dependence on the effective gate voltage is shown in the inset of figure\,\ref{alpha}(b). 
	
	Summarizing the experimental data, we conclude that the observed photoresponse from graphene in the quantum Hall regime stems from the edges channels and the direction of the photocurrent along an edge is defined solely by the magnetic field polarity. 
	
	\section{Discussion}
	Now we discuss the microscopic origin of edge photocurrents. 
	Edge photocurrents at zero magnetic field have been studied previously in low mobility graphene samples~\cite{edge}. 
	Under this conditions, the $dc$ current formation involves the 
	motion of the charge carriers under the action of the $ac$ electric field and the scattering at the sample edge. 
	The semiclassical theory of the edge photogalvanic effect~\cite{GlazovGanichev_review,edge} shows that the $dc$ electric current is formed in the vicinity of the sample edge within the width of about the carrier mean free path. 
	The current direction is determined by the radiation polarization and is opposite for samples with electron and hole gases. 
	In the case of linearly polarized radiation, the polarization dependence follows $\sin(2\alpha')$.
	Exactly this behavior is detected in our experiments on high mobility samples at zero magnetic field, see figures\,\ref{alpha}(a)-(c).
	
\begin{figure}
		\includegraphics[width=0.95\linewidth]{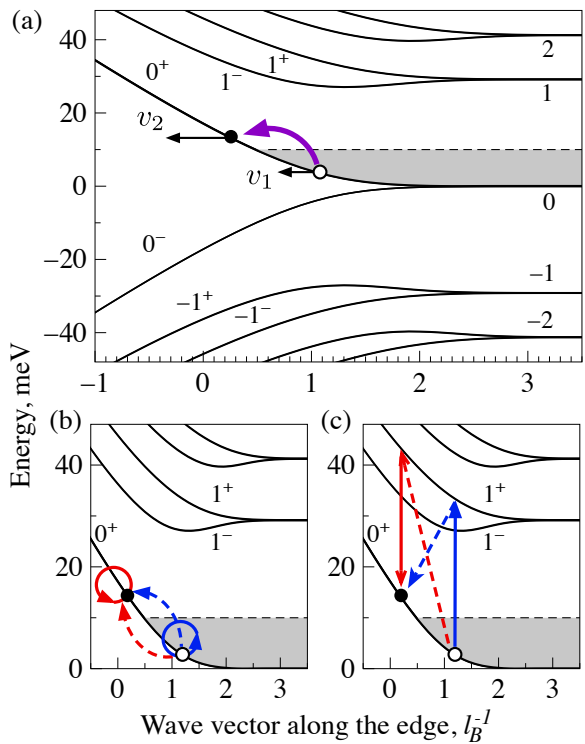}
		\caption{
			(a) Schematic illustration of the terahertz radiation induced edge current in graphene quantum Hall insulator. 
			Panel shows the calculated spectrum of Landau levels in graphene with an armchair edge. 
			The energy is given in absolute units corresponding to $B = 1$~T, the dashed line illustrates the position of the Fermi level for small positive effective gate voltages resulting in $n$-type conductivity. 
			The photocurrent emerges as a result of electron transitions between states with different velocities under absorption of a photon (thick arrow). 
			(b)-(c) Zooms of electron dispersion with illustration of possible mechanisms of intraband photon absorption. 
			The solid arrows illustrate interaction with the electro-magnetic field, whereas the dashed arrows stand for scattering by static impurities or phonons.
		}
		\label{model} 
	\end{figure}

	The above mechanism, however, fails for the quantum Hall regime, for which the Fermi level lies in the gap between the Landau levels, the carriers are localized, and the sample conductivity is determined by one-dimensional edge channels. 
	To uncover the mechanism of the photocurrent generation, we start with the analysis of the energy dispersion of graphene in the quantum Hall regime. 
	Figure\,\ref{model}(a) shows the energy spectrum of a semi-infinite graphene sheet in the field $B = 1$~T calculated in the microscopic model of \cite{Levitov2007} for the exemplary armchair edge. 
	The bulk Landau levels are two-fold degenerate in the valley index and have the energies $\eps_n \propto \sqrt{|n|}~\mathrm{sgn}\,n$, where $n$ is an integer number. 
	In the figure\,\ref{model}(a) this corresponds to $k l_B \gg 1$, where $k$ is the wave vector along the edge defining also the position of the electron orbit in real space and $l_B$ is the magnetic length. 
	At the edge ($k l_B \lesssim 1$), the atomically sharp potential couples the Landau levels from the $K$ and $K'$ valleys, which leads to the formation of ``valley-symmetric'' and ``valley-antisymmetric'' edge states denoted by $n^+$ and $n^-$, respectively, in figure\,\ref{model}(a). 
	The energy spectrum mirrors with respect to zero energy providing that the edge potential preserves the particle-hole symmetry.
	We note that the Zeeman splitting in graphene is much smaller than the cyclotron energies for the studied range of magnetic fields and is therefore neglected.
	%

	Consider now the excitation of the graphene structure by terahertz radiation, see figure\,\ref{model}(a). 
	This sketch corresponds to the experimentally relevant situation of the quantum Hall plateau with the filling factor $\nu = 2$ (due to the account for spin degeneracy) achieved for $|B| \geq 0.5$~T, see figure\,\ref{transport}(a). 
	In this case, the Fermi level lies between the $n = 0$ and $n = 1$ orbital bulk Landau levels and the energy gap between the Landau levels exceeds the photon energy used in the experiments.
	The absorption of photons occurs then in the edge channels as the result of indirect (Drude-like, where the momentum conservation is 
	satisfied by the interaction with static impurities or phonons) optical transitions, as shown in figure\,\ref{model}(a).
	In these processes, electrons in the initial states with the velocity $v_1$ below the Fermi level make transitions to the final states with the velocity $v_2$ above the Fermi level. 
	The difference in the velocities $v_1$ and $v_2$ stemming from the dispersion of the edge states results in the formation of a $dc$ electric current along the sample edge.
	In the relaxation time approximation, the edge photocurrent is given by
	\begin{align}
	\label{jedge}
	j = q \frac{4 \pi}{\hbar} \sum \limits_{k, k'} \left[ \tau_{\rm edge}(\eps_{k'}) v_{k'} - \tau_{\rm edge}(\eps_{k}) v_{k} \right] \times \\ 
	\times |M_{k' k}|^2 (f_{k} - f_{k'}) \delta (\eps_{k'} - \eps_k - \hbar \omega), \nonumber
	\end{align}
	where $q$ is the carrier charge ($q < 0$ for the $n$-type conductivity), $k$ ($k'$) is the wave vector of the initial (final) state, $\eps_{k}$ and $v_k = (1/\hbar) d \eps_{k}/dk$ are the energy and the group velocity, respectively, $M_{k' k}$ is the matrix element of the indirect intralevel transitions, $f_{k}$ is the equilibrium distribution function, and $\tau_{\rm edge}$ is the relaxation time of electron excitations in the chiral edge channels. 
	As follows from \eqref{jedge}, the edge photocurrent emerges if the products $\tau_{\rm edge} v$ are different for the initial and final states. 
	The general expression for $j$ can be simplified if $\hbar \omega \ll \eps_{F}$, where $\hbar \omega$ is the photon energy and $\eps_{F}$ is the Fermi energy counted from the zero Landau level in the bulk. 
	In this case, the edge photocurrent can be presented in the form
	\begin{align}
	\label{jedge2}
	j = q \left[ v_{F} \frac{\partial \tau_{\rm edge}}{\partial \eps} (\eps_{F}) + \tau_{\rm edge}(\eps_{F}) \frac{\partial v}{\partial \eps} (\eps_{F}) \right] w I\:,
	\end{align}
	where $v_F$ is the edge-state velocity at the Fermi level, $w$ is the absorption width of the channel, and $I$ is the radiation intensity. 
	The product $w I$ is the power absorbed by the edge of the unit length.
	
	Equation~\eqref{jedge2} describes the main features of the edge photocurrent observed in the experiment. 
	First, it shows that the photocurrent direction is determined by the magnetic field and is reversed when the magnetic field sign is switched. 
	Indeed, the replacement $B \to - B$ results in the reverse of the edge-state velocity $v$ and its derivative $\partial v/\partial \eps$, while $\tau_{\rm edge}$ and its derivative $\partial \tau_{\rm edge}/\partial \eps$ remain unchanged. 
	Hence, the photocurrent reverses its direction. 
	This is in agreement with figures\,\ref{alpha}(d) and \ref{alpha-gate-B}, which show that the current directions are opposite for positive and negative $B$. 
	
	Second, for a fixed magnetic field polarity, the edge photocurrents given by \eqref{jedge2} in $n$- and $p$-type structures flow in the same direction. 
	Figure\,\ref{model}(a) illustrates the mechanism of the current generation for the filling factor $\nu = 2$ in the electron system which is realized for a positive effective gate voltage. 
	At a negative effective gate voltage, see figure\,\ref{transport}, we observe the quantum Hall plateau with the hole filling factor $\nu = 2$, which corresponds to the Fermi level lying between the $n = -1$ and $n = 0$ bulk Landau levels (in the electron representation). 
	In this case, the photocurrent is formed by hole excitations in the hole edge channels. 
	Importantly, the charge photocurrent $j$ is excited in the same direction as for the electron case because both the carrier charge $q$ and the edge-state velocity $v_k \propto \partial \eps_k / \partial k$ 
	reverse their signs. 
	Such an invariance of the edge photocurrent with respect to charge conjugation is also in agreement with the experiment. 
	The data for different gate voltages shown in figure\,\ref{alpha-gate-B} demonstrate that the photocurrent direction does not depend, indeed, on the type of carriers. 
	
	The observed polarization dependencies see figures\,\ref{alpha} and\,\ref{alpha-gate-B}, are determined by the polarization dependence of the edge-channel absorption width $w$ in \eqref{jedge2}. 
	We calculate it for the indirect intralevel optical transitions sketched in figure\,\ref{model}(a). 
	These are second-order processes which require the simultaneous electron-photon interaction and electron scattering by static impurities or phonons to satisfy the energy and momentum conservation~\cite{Tarasenko2007}. 
	The indirect transitions are theoretically described by virtual processes through intermediate states which can belong either to the same level or to other levels in the conduction or valence bands. 
	Figures\,\ref{model}(b) and (c) illustrate such virtual transitions with the solid and dashed curves denoting the electron-photon interaction and the electron scattering, respectively. 
	The virtual transitions with intermediate states in the same level, figure\,\ref{model}(b), are induced by radiation polarized along the edge whereas other virtual transitions, see figure\,\ref{model}(c) as an example, can be induced by both parallel and normal to the edge components of the radiation electric field. 
	Such a polarization sensitivity of the optical transitions for linearly polarized radiation leads to the linear dichroism where the polarization dependence of the absorption width is given by 
	\begin{equation}\label{eta}
	w = w_{\parallel} |e_{\parallel}|^2 + w_{\perp} |e_{\perp}|^2 = \frac{w_{\parallel} + w_{\perp}}{2} + \frac{w_{\parallel} - w_{\perp}}{2} \cos 2 \alpha'\:,
	\end{equation}
	where $w_{\parallel}$ and $w_{\perp}$ are the absorption widths for the radiation polarized along and normal to the edge, respectively, and $\bm{e}$ is the unit vector of the radiation polarization. 
	At $\hbar\omega \ll \eps_F$, the transitions excited by the electric field parallel to the sample edge dominate over the transitions excited by the electric field normal to the edge, and hence $w_{\parallel} \gg w_{\perp}$.
	The polarization dependence of the edge photocurrent given by \eqref{jedge2} and~\eqref{eta} is shown in figure\,\ref{alpha}(b) by dashed lines and fits reasonably well the experimental data. 
	The experimentally observed phase shift, which is small for low densities and radiation frequencies, and more pronounced at high densities and frequencies, see figure\,\ref{alpha-gate-B}, may be caused by electron gas heating. 
	We also note that, for elliptically or circularly polarized radiation, the optical transitions are sensitive to the sign of circular polarization. 
	Therefore, we expect the circular dichroism in the edge channels and the contribution to the edge photocurrent sensitive to the photon helicity.
	
	Finally, we compare the photocurrent magnitude detected in the experiment with that expected from the microscopic model. 
	The absorption width of a chiral edge channel for the indirect optical transitions accompanied by electron scattering from randomly distributed defects can be estimated as $w \sim \alpha l_B /(\sqrt{n_s} \,v_0 \tau_p)$, where $\alpha$ is the fine-structure constant, 
	$l_B = \sqrt{\hbar c/|q B|}$, $v_0$ and $\tau_p$ are the velocity and momentum relaxation of electrons 
	in graphene at zero magnetic field, respectively, and $n_s$ is the electron density. This gives a theoretical estimation for the edge photocurrent magnitude
	\begin{equation}
	\label{amplitude}
	\left| j/I \right| \sim \frac{\left| q \right| \alpha \, l_B^2}{\hbar v_0 \sqrt{n_s}} \left(\frac{\tau_{\rm edge}}{\tau_p}\right) \:.
	\end{equation}
	Equation~\eqref{amplitude} suggests an overall decrease of the photocurrent magnitude with the magnetic field since $j \propto l_B^2 \propto 1/B$.
	Such a decrease is observed in the experiment, see the data in the inset in figure\,\ref{alpha-gate-B}(c) for the magnetic field range of the 
	quantum Hall regime which is highlighted by gray background~\cite{footnoteclassic}
	The theoretical estimation fits to the measured amplitude $J/I \sim 0.2$~nA$\,$cm$^2$/W, see figure\,\ref{alpha}(b), for $\tau_{\rm edge}/\tau_p \sim 10^2$. 
	The large ratio $\tau_{\rm edge}/\tau_p \sim 10^2$ is not surprising because $\tau_{\rm edge}$ is determined by the processes of electron energy relaxation in uni-direction chiral channels while $\tau_p$ is limited by elastic scattering. 
	The relaxation of electron excitations in chiral channels can be caused by the interaction with acoustic phonons. However, the standard single-phonon processes are suppressed because of the drastic difference in phonon and electron velocities. The relaxation can be more efficient in two-phonon processes or supercollisions~\cite{Song2012}, where electrons are simultaneously scattered by phonons and static defects. 
	Another possible mechanism of the relaxation involves the tunneling of carriers between the edge states and the states localized in the bulk of graphene near its boundary. 
	The microscopic calculation of $\tau_{\rm edge}$ is an interesting task for future research.

	\section{Conclusion}
	We reported on the observation and study of THz induced 
	edge currents in high mobility exfoliated graphene in the quantum Hall effect regime. 
	The edge current is generated by Drude-like absorption of THz photons at the quantum Hall edge states resulting in a net velocity of the charge
	carriers. 
	Unlike the case of zero magnetic field, the direction of edge photocurrents is only determined by the sign of the magnetic field and not by the charge carrier type. 
	All experimental findings, in particular the dependence on linear polarization, are explained with the microscopic model developed here. 
	We also determine the edge scattering time to be about two orders of magnitude longer than the momentum relaxation time of bulk carriers at zero magnetic field.
	
	\section*{Acknowledgements} 
	We thank D. Kozlov for discussions. 
	The support from the DFG priority program SFB 1277 (project A04) and GRK1570, and the Elite Network of Bavaria (K-NW-2013-247) is gratefully acknowledged.
	S.A.T. received support from the Government of the Russian Federation (contract \# 14.W03.3.0011 at Ioffe Institute). 
	M.V.D. acknowledges financial support from the RFBR (project No.16-32-60175).

\end{document}